\documentclass[
english,			
french,				
spanish,			
brazil,				
]{article}

\usepackage{arxiv}
\usepackage[utf8]{inputenc} 
\usepackage[T1]{fontenc}    
\usepackage{hyperref}       
\usepackage{url}            
\usepackage{booktabs}       
\usepackage{amsfonts}       
\usepackage{nicefrac}       
\usepackage{microtype}      
\usepackage{lipsum}		
\usepackage{graphicx}
\usepackage{amssymb}
\usepackage{amstext}
\usepackage{amsmath}
\usepackage{amsthm}
\usepackage{subfig}  
\usepackage{stfloats}   
\usepackage{textcomp}

\newcommand{\rvx}{\textbf{\emph{x}}}

\newcommand{\vax}{\textbf{\emph{x}}}
\newcommand{\vay}{\textbf{\emph{y}}}

\newcommand{\var}{\textbf{\emph{r}}}

\title{Covid 19 and A Wavelet Analysis of the Total Deaths per Month in Brazil since 2015}

\author{ \href{https://orcid.org/0000-0003-3557-1957}{\includegraphics[scale=0.06]{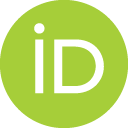}\hspace{1mm}Alexandre Barbosa de Lima} \\
	Member, IEEE \\
	School of Engineering,  University of S\~ao Paulo \\
	Department of Energy and Automation -- Researcher \\
	\texttt{alexandreblima@usp.br} 
}

\hypersetup{
pdftitle={A template for the arxiv style},
pdfsubject={q-bio.NC, q-bio.QM},
pdfauthor={David S.~Hippocampus, Elias D.~Striatum},
pdfkeywords={First keyword, Second keyword, More},
}

\begin{document}
\maketitle

\begin{abstract}
We investigate the historical series of the total number of deaths per month in Brazil since 2015 using the wavelet transform, in order to assess whether the COVID-19 pandemic caused any change point in that series. Our wavelet analysis shows that the series has a change point in the variance. However, it occurred long before the pandemic began.
\end{abstract}


\section{Introduction}\label{sec:intro}

Until the time of this writing, the Brazilian Federal Ministry of Health has recorded 3,846,153 Sars-CoV-2 case reports and $120,462$ deaths caused by the COVID-19 pandemic \cite{minsaude} , \cite{WHO}.

According to the Coronavirus Resource Center of the Johns Hopkins University (JHU) \cite{JHU}, Brazil is the second country most affected by COVID-19 in the world, both in number of cases and in number of deaths.

This article aims to investigate the historical series of the total number of deaths using the wavelet transform, which is a powerful time-frequency domain signal processing  tool \cite{percival00}. From now on, such series will be referred simply as \textit{historical series} or \textit{signal} .

Wavelet analysis is capable of solving problems, such as the detection of the \textit{change point} in a signal (a transient phenomena \cite{chen2000}, \cite{killick2013}), which are difficult to approach with tools such as the Windowed Fourier Transform (WFT) \cite{lima2013}. A signal is any sequence of observations associated with an ordered independent variable $t$, which can be discrete or continuous \cite{percival00}.

The signal is interrogated using wavelet analysis in order to answer the following questions: i) Does the series present any change point(s)? ii) If there is a change point, what is its type (change in  level, variance, etc. )? iii) If there is a change point, did it occur during the COVID-19 pandemic?

The database used is that made available online by the Transparency Portal of the Civil Registry Offices of Brazil \cite{database}, which consolidates the amount of birth, marriage and death certificates available in Brazil. Online data has been available since January 2015.
Brazilian registries are regulated by the National Council of Justice (CNJ), which is a public institution headquartered in Brasília, Federal District, that aims to improve the work of the Brazilian judicial system, especially with regard to administrative and procedural control and transparency \cite{cnj}. The president of the Brazilian Supreme Court also presides the CNJ.

The analysis was performed using the \texttt{R} software, version 4.0.2 \cite{R} and the \texttt{MATLAB}\textcopyright$\,$R2015a Wavelet Toolbox. The \texttt{R} and \texttt{MATLAB}\textcopyright$\,$codes, as well as the database in Excel spreadsheet format, are available for public consultation  and auditing on \texttt{GitHub} \cite{GitHub}.

The remainder of the paper is organized as follows.  Section \ref{sec:wavelets} presents an overview of the Continuous and Discrete Wavelet Transforms for the reader who is not familiar with the subject. If the reader is familiar with the theory of wavelets, he or she may proceed to Section \ref{sec:results}, which presents the experimental results. Finally, section \ref{sec:conclusions} presents our conclusions and and highlights some topics for further work.

\section{Wavelets}\label{sec:wavelets}

\subsection{The Continuous Wavelet Transform}\label{subsec:CWT}

The Fourier Transform (FT) of a signal $x(t)$, $t \in \mathbb{R}$ ($t$ denotes time), if exists, is defined as 

\begin{equation}
\label{def:TF}
X(\nu) = \text{TF}\{x(t)\} = \int_{-\infty}^{\infty} x(t) e^{-j2\pi\nu t}dt\,
\end{equation}
in which $\nu$ denotes the frequency in cycles/second [Hz].

Gabor \cite{gabor46} has shown that it is possible to represent the local spectral content of a signal $x(t)$ around an instant of time  $\tau$ by the Windowed Fourier Transform (WFT).

\begin{equation}
\label{def:WFT}
X_T(\nu,\tau) = \int_{-\infty}^{\infty} x(t)g_T(t-\tau)e^{-j2\pi\nu t}dt \, ,
\end{equation}

in which $g_T(t)$ is a window of finite duration support $T$ and $\nu$ denotes frequency.

The WFT is a two-dimensional representation defined on the time-frequency  domain (or plane) as it depends on the $\nu$ and $\tau$ parameters. The WFT would be equivalent to a kind of continuous ``sheet music'' description of $x(t)$.  

According to the Heisenberg's uncertainty principle \cite[p.52]{kaiser94}, a signal whose energy content is quite well localized in time has this energy quite spread out in the frequency domain. As the window of (\ref{def:WFT}) has a fixed size $T$, we may conclude that the WFT is not good to analyze (or identify) behaviors of $x(t)$ occurring in time intervals much smaller or much larger than $T$, as, for example, transient phenomena of duration $\Delta t << T$ or cycles that exist in periods larger than  $T$. 

A  \emph{wavelet} $\psi_0(t)$ (sometimes also called mother wavelet), $t \in \mathbb{R}$, is a function that satisfies three conditions \cite{percival00}, \cite{whitcher2001}.

\begin{enumerate}
	\item \label{admissib}
	Its Fourier transform $\Psi(\nu)$, $-\infty < \nu < \infty$, is such that exists a finite constant
	$C_\psi$ that obeys the \textit{admissibility condition}
	\begin{equation}
	\label{eq:admissib}
	0< C_\psi = \int_{0}^{\infty} \frac {|\Psi(\nu)|^2}{\nu} d\nu < \infty \,.
	\end{equation}
	\item \label{cruza-zeros}
	The integral of $\psi_0(t)$ is null:
	\begin{equation}
	\label{eq:cruza-zeros}
	\int_{-\infty}^{\infty} \psi_0(t) \,dt = 0\,.
	\end{equation}
	\item \label{sup-finito}
	Its energy is unitary:
	\begin{equation}
	\label{eq:sup-finito}
	\int_{-\infty}^{\infty} |\psi_0(t)|^2 \,dt = 1\,.
	\end{equation}						
\end{enumerate}

\begin{figure}[htp]
	\begin{center}
		\includegraphics[scale=0.75]{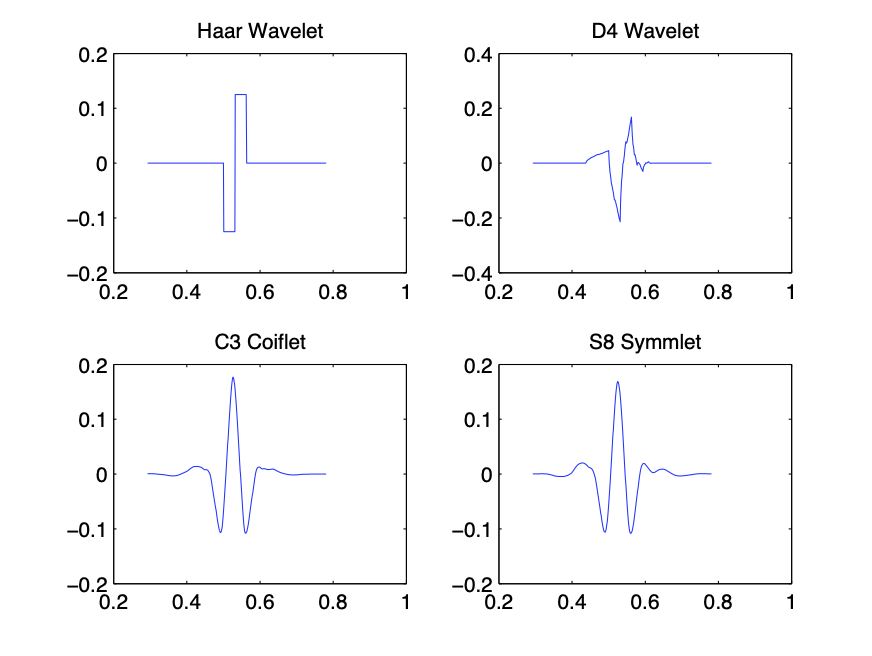}
	\end{center}
	\caption{four examples of wavelet functions.}
	\label{fig:examples-of-wavelets}
\end{figure}

Figure \ref{fig:examples-of-wavelets} shows four examples of wavelet functions: Haar, Daubechies, Coiflet and Symmlet. As the name suggests, a wavelet is a `small wave'. A small wave grows and decays in a limited time period. On the other hand, an example of a `big wave' is the cosine function $\cos t$, which is `eternal', i. e., keeps oscilating up and down for all $t$.

The wavelet transform is a relatively new tool for the analysis of signals, given that his mathematical theory was formalized in the 1980s \cite{daubechies88}. The wavelet transform has been originally developed as an analysis and synthesis tool of continuous time energy signals \cite{morlet84}, \cite{mallat89}, \cite{mallat89b}, \cite{mallat89c}, \cite{daubechies88}, \cite{daubechies92}.

An energy signal $x(t)$, obeys the constraint

\begin{equation}
\label{eq:sinal-de-energia}
\left\|x\right\|^2 = \left\langle x,x\right\rangle \equiv \int_{-\infty}^{\infty} |x(t)|^2\,dt < \infty,
\end{equation}
i. e.,  $x(t)$ that obeys the constraint (\ref{eq:sinal-de-energia}) belongs to the squared summable functions space $L^2(\mathbb{R})$. 

Presently, the wavelet transform has also been used as an analysis tool of discrete time signals.

There are continuous time and discrete time wavelet decompositions designate by Continuous Wavelet Transform (CWT) and Discrete Wavelet Transform (DWT).

The CWT of a signal $x(t)$ consists of a set $C=\{W_\psi(s,\tau), \,s \in \mathbb{R}^+, \,\tau \in \mathbb{R} \}$, in which

\begin{itemize}
	\item $\tau$ is the time localization parameter,
	\item $s$ represents scale, and 
	\item $\psi$ denotes a wavelet function,
\end{itemize}

of wavelet coefficients on the continuous time-scale plane (also known as time-frequency plane) given by 

\begin{equation}
\label{eq:CWT}
W_\psi(s,\tau) = \left\langle \psi_{0_{(s,\tau)}},x \right\rangle =
\int_{-\infty}^{\infty} \frac{1}{\sqrt{s}} \psi_0^{\ast}\left( \frac{\lambda-\tau}{s} \right) x(\lambda) d\lambda 
\end{equation}

$\psi_{0_{(s,\tau)}}(t)=s^{-1/2} \psi_0\left( \frac{t-\tau}{s} \right)$ denotes a dilated and shifted version of the ``mother'' wavelet  $\psi_0(t)$. 

The factor $1/\sqrt{s}$ in (\ref{eq:CWT}) provides all functions of the class 

\begin{equation}
\label{eq:psi-analise}
\mathcal{W} = \left\{  \frac{1}{\sqrt{s}} \psi_0\left( \frac{t-\tau}{s} \right) \in \mathbb{R} \right\} \, 
\end{equation}

have the same energy (norm).

The basic idea of the CWT defined by (\ref{eq:CWT}) is to correlate\footnote{Measure the similarity.} a signal $x(t)$ with shifted (by $\tau$) and dilated (by $s$) versions of a mother wavelet (that has a pass-band spectrum). The CWT is a two parameters function. So, it is a redundant transform, because it consists on mapping an one-dimension signal on the time-scale plane.  

Differently from the  WFT, where the reconstruction is made from the same family of functions as that used in the analysis, in the CWT the synthesis is made with functions $\tilde{\psi}_{s,\tau}$ that have to satisfy

\begin{equation}
\label{eq:psi-sintese}
\tilde{\psi}_{s,\tau}(t) = \frac{1}{C_\psi}\frac{1}{s^2} \psi_{s,\tau}(t) \,.
\end{equation}  

So, $x(t)$ is completely recovered by the Inverse Continuous Wavelet Transform (ICWT):

\begin{equation}
\label{eq:ICWT}
x(t) = \frac{1}{C_\psi} \int_{0}^{\infty} \left[ \int_{-\infty}^{\infty} W_\psi(s,\tau)	\frac{1}{\sqrt{s}} \psi\left( \frac{t-\tau}{s} \right) d\tau\right] \frac{ds}{s^2}\,.
\end{equation}

The fundamental difference between the CWT and the WFT consists of the fact that the functions $\psi_{s,\tau}$ undergo dilations and compressions \cite{kaiser94}. The analysis on refined scales of time (small values of $s$) requires ``fast'' $\psi_{s,\tau}$ functions, i. e., of a small support, while the analysis on aggregate scales of time (large values of $s$) requires ``slower'' $\psi_{s,\tau}$ functions, i. e., of a wider support. As already mentioned, the internal product defined by (\ref{eq:CWT}) is a measure of similarity between the wavelet $\psi\left( \frac{t-\tau}{s} \right)$ and the signal $x(t)$ on a certain instant of time $\tau$ and on a determined scale $s$. For a fixed $\tau$, large values of $s$ correspond to a low-frequency analysis, while small values of $s$ are associated to a high-frequency analysis. Therefore, the wavelet transform has a \textit{variable time resolution} (i. e., the capacity of analyzing a signal from close - ``\emph{zoom in}'' - or from far - ``\emph{zoom out}''), being adequate to analyze phenomena that occur in different time scales.

\begin{figure}[htp]
	\begin{center}
		\includegraphics[scale = 0.5]{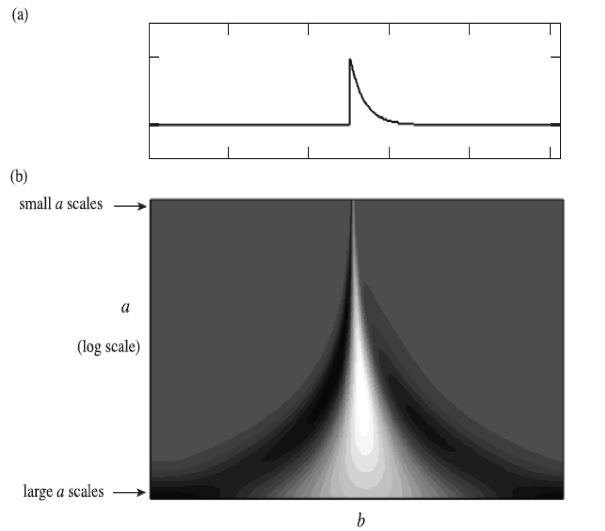}
	\end{center}
	\caption{Analysis of an exponential discontinuity. The image on the bottom part of the figure is the CWT $W_\psi(s,\tau)$ of the signal on the top part. The parameters $\tau=b$ and $s=a$ vary along the horizontal and vertical axes, respectively. Large (positive) values of wavelet coefficients are indicated in white. Note that the discontinuity produces large coefficients in its respective cone of influence, which converge to the location of the singularity.}
	\label{fig:cwt-exponencial}
\end{figure}

Figure \ref{fig:cwt-exponencial} provides an example of a CWT.

\subsection{Multiresolution Analysis and the Discrete Wavelet Transform}\label{subsec:mra}

There are two kinds of DWT:

\begin{itemize}
	\item the DWT for discrete time signals; and
	\item the DWT for continuous time signals.
\end{itemize}

The DWT may be formulated for discrete time signals (as it is done, for example, by Percival and Walden \cite{percival00}) without establishing any explicit connection with the CWT. On the other hand, we should not understand the term ``discrete'' of the DWT for continuous time signals as meaning that this transform is defined over a discrete time signal. But only that the coefficients produced by this transform belong to a subset $D=\{w_{j,k} = W_\psi(2^{j},2^jk), \,j \in \mathbb{Z}, \,k \in \mathbb{Z} \}$ of the set $C$ \cite[p.105]{whitcher2001}, \cite{veitch00b}. 

In fact, the DWT coefficients for continuous time signals can also be directly obtained by means of the integral

\begin{equation}
\label{eq:integral-DWT}
w_{j,k} = \left\langle \psi_{0_{(2^j,2^jk)}},x \right\rangle = \int_{-\infty}^{\infty} 2^{-j/2} \psi_0^{\ast}(2^{-j}\lambda - k) x(\lambda) \,d\lambda \,,
\end{equation}

in which the indices $j$ and $k$ are called scale and localization, respectively, does not involve any discrete time signal, but the continuous time signal  $x(t)$. 

Equation (\ref{eq:integral-DWT}) shows that the continuous time DWT corresponds to a critically sampled version of the CWT defined by (\ref{eq:CWT}) in the dyadic scales $s=2^{j}$, $j=\ldots, -1, 0, 1, 2,\ldots$, in which the instants of time in the dyadic scale $s=2^j$ are separated by multiples of $2^j$. The function $\psi_0$ of (\ref{eq:integral-DWT}) must be defined from a Multiresolution Analysis (MRA) of the signal $x(t)$ \cite{percival00}, \cite{daubechies92}, \cite{mallat99}. Observe that the continuous time MRA theory is similar to that of discrete time.

In this paper, we decided, for mere convenience, to present the continuous time MRA version based on the spectral analysis of a ``fictitious''  signal $\{\tilde{\rvx}_t,\, t \in \mathbb{R}\}$ that is associated to the discrete time series  $\{\rvx_n,\, n \in \mathbb{Z}\}$ \cite{veitch00b}.

\begin{figure}[htp]
	\begin{center}
		\includegraphics[scale=0.5]{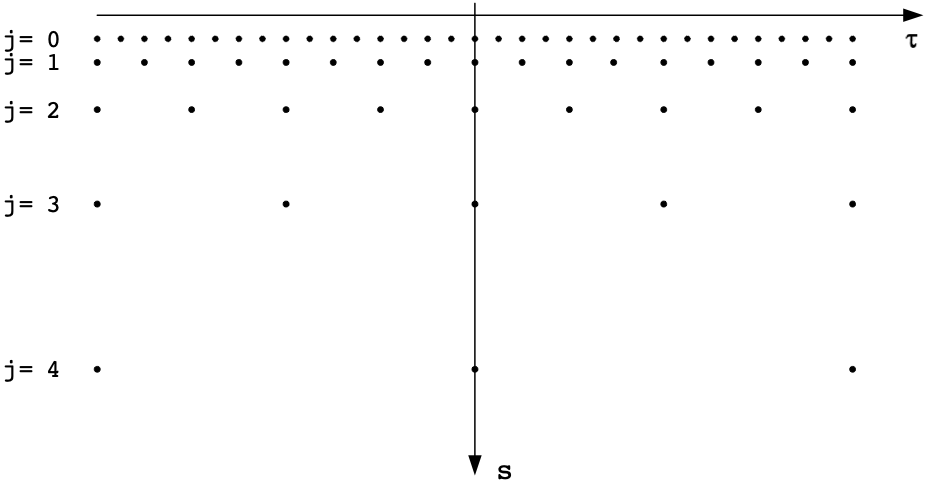}
	\end{center}
	\caption{Critical sampling of the time-scale plane by means of the CWT parameters ($s=2^j$ e $\tau=2^jk$) discretization. }
	\label{fig:cwt-dwt}
\end{figure}

Figure \ref{fig:cwt-dwt} shows the critical sampling of the time-scale plane by means of the CWT parameters ($s=2^j$ e $\tau=2^jk$) discretization. 

A MRA is, by definition, a sequence of closed subspaces $\{V_j\}_{j \in \mathbb{Z}}$ of $L^2(\mathbb{R})$  such that \cite[p.462]{percival00}, \cite{daubechies92}:

\begin{enumerate}
	\item $\ldots V_2 \subset V_1 \subset V_0 \subset V_{-1} \subset V_{-2} \subset \ldots$;
	\label{subconjuntosVj}
	\item $\bigcap_{j \in Z} V_j = \{\}$; 
	
	\item $\bigcup_{j \in Z} V_j = L^2(\mathbb{R})$;
	\label{Vj_forma_L2}
	\item $x(t) \in V_j \Leftrightarrow x(2^j t) \in V_0, j>0$ (in which $t$ denotes time and $x(t)$ is an energy signal);
	\label{fdesloc}
	\item There is a function $\phi_{j}(t) = 2^{-j/2}\phi_0(2^{-j}t)$ in $V_j$, called \textit{scale function}, such that the set $\{\phi_{j,k}, \: k \in \mathbb{Z}\}$ is an orthonormal basis of $V_j$, with $\phi_{j,k}(t) = 2^{-j/2} \phi_0(2^{-j}t - k) \: \forall j,k \in \mathbb{Z}$.
	\label{subespacophi}
\end{enumerate}

The subspace $V_j$ is known as the \textit{approximation space} associated to the time scale $s_j=2^j$ (assuming that $V_0$ is the approximation space with unit scale).

If the $x(t)$ projection on $V_j$  is represented by the scale coefficients

\begin{equation}
\label{eq:int-coef-escala}
u_{j,k}=\left\langle \phi_{j,k},x\right\rangle = \int_{-\infty}^{\infty} 2^{-j/2} \phi_0^{\ast}(2^{-j}t - k)x(t)\, dt, 
\end{equation}

then properties \ref{subconjuntosVj} and \ref{Vj_forma_L2} assure that $\underset{j\to -\infty}\lim \sum_k \phi_{j,k}(t) u_{j,k} = x(t)$, $\forall\,\,x \in L^2(\mathbb{R})$. 

Property  \ref{fdesloc} implies that the subspace $V_j$ is a scaled version of subspace $V_0$ (multiresolution).

The orthonormal basis mentioned in property \ref{subespacophi} is obtained by time shifts of the low-pass function $\phi_{j}$.

Consider the successive approximations sequence (also known in the literature as \emph{wavelet smooths} \cite{percival00}) of $x(t)$ 
\begin{equation}
\label{eq:aprox}	
\mathcal{S}_j(t) = \sum_k \phi_{j,k}(t) u_{j,k} \quad j=\ldots,-1,0,1,\ldots \,.
\end{equation}

As $V_{j+1} \subset V_{j}$, $\mathcal{S}_{j+1}(t)$ is a coarser approximation of $x(t)$ than $\mathcal{S}_{j}(t)$. 

This fact illustrates the MRA's fundamental idea, that consists in examining the \textit{loss of information} when one goes from $\mathcal{S}_{j}(t)$ to $\mathcal{S}_{j+1}(t)$:

\begin{equation}
\label{eq:detalhe}	
\mathcal{S}_{j}(t) = \mathcal{S}_{j+1}(t) + \Delta x_{j+1}(t).
\end{equation}

$\Delta x_{j+1}(t)$ (called \textit{detail} of $x_j(t)$) belongs to the subspace $W_{j+1}$, named detail space \cite{percival00} that is associated to the fluctuations (or variations) of the signal in the more refined time scale $s_j=2^j$  and that corresponds to the orthogonal complement of $V_{j+1}$ in  $V_{j}$\footnote{Besides, $W_{j+1}$ is contained in the subspace $V_{j}$.}.

The MRA shows that the detail signals $\Delta x_{j+1}(t) = \mathcal{D}_{j+1}(t)$ may be directly obtained by successive projections of the original signal $x(t)$ over wavelet  subspaces $W_j$. 

Besides, the MRA theory shows that exists a function $\psi_0(t)$, called ``mother wavelet'' , that is obtained from $\phi_0(t)$, in which $\psi_{j,k}(t) = 2^{-j/2} \phi_0(2^{-j}t - k) \:, k \in \mathbb{Z},$ is an orthonormal basis of $W_j$.    

The detail $\mathcal{D}_{j+1}(t)$ is obtained by the equation
\begin{equation}
\label{eq:detalhe2}	
\mathcal{D}_{j+1}(t) = \sum_k \psi_{j+1,k}(t)\left\langle \psi_{j+1,k}(t), x(t)\right\rangle \,. 
\end{equation}

The internal product $\left\langle \psi_{j+1,k}(t), x(t)\right\rangle = w_{j+1,k}$ denotes the wavelet coefficient associated to scale $j+1$ and discrete time $k$ and $\{\psi_{j+1,k}(t)\}$ is a family of wavelet functions that generates the subspace $W_{j+1}$, orthogonal to subspace $V_{j+1}$ ($W_{j+1} \bot V_{j+1}$), i. e., 
\begin{equation}
\label{eq:<phi,psi>}
\left\langle \psi_{j+1,n} , \phi_{j+1,p} \right\rangle = 0 \, , \forall n,p.
\end{equation}

Therefore, the detail signal $\mathcal{D}_{j+1}(t)$ belongs to the complementary subspace $W_{j+1}$ of $V_{j}$, because 

\begin{equation}
\label{eq:soma-direta}
V_{j} = V_{j+1} \oplus W_{j+1}.
\end{equation}

That is, $V_{j}$ is given by the direct addition of $V_{j+1}$ and $W_{j+1}$, and this means that any element in $V_{j}$ may be determined from the addition of two orthogonal elements belonging to $V_{j+1}$ and $W_{j+1}$. Iterating (\ref{eq:soma-direta}), we have
\begin{equation}
\label{eq2:soma-direta}
V_{j} = W_{j+1} \oplus W_{j+2} \oplus \ldots \quad.
\end{equation}

Eq. (\ref{eq2:soma-direta}) says that the approximation $\mathcal{S}_j(t)$ is given by
\begin{equation}
\label{eq:sintese-aproxj}	
\mathcal{S}_j(t)  =  \sum_{i=j+1}^{\infty}\sum_{k} w_{i,k} \psi_{i,k}(t)\,. 
\end{equation}

The MRA of a continuous time signal $x(t)$ is initiated by determining the coefficients\footnote{The sequence
	$u_0(k)$ is obtained sampling the filter's output whose impulse response is $\phi^{\ast}(-t)$ (matched filter with a function $\phi_0(t)=\phi(t)$) at instants $k=0,1,2,\ldots$, i. e., $u_0(k) = x(t)\star \phi^{\ast}(-t)$ for $k=0,1,2,\ldots$, in which $\star$ denotes convolution.}
$u_0(k)=\left\langle \phi_{0,k}(t),x(t)\right\rangle$, in which $k=0,1,\ldots,N-1$,
that are associated to the projection of $x(t)$ on the approximation subspace $V_0$. 

Following, the sequence $\{u_0(k)\}$ is decomposed by filtering and sub-sampling by a factor of $2$ (\emph{downsampling}) in two sequences: $\{u_1(k)\}$ and $\{w_1(k)\}$, each one with $N/2$ points. This filtering and sub-sampling process is repeated several times, producing the sequences

\begin{equation}
\label{eq:seq-u}
\{ \{u_0(k)\}_N,\{u_1(k)\}_{\frac{N}{2}},\{u_2(k)\}_{\frac{N}{4}},\ldots, \{u_j(k)\}_{\frac{N}{2^j}},\ldots,\{u_{J}(k)\}_{\frac{N}{2^{J}}}\}
\end{equation} 
and 
\begin{equation}
\label{eq:seq-w}
\{ \{w_1(k)\}_{\frac{N}{2}},\{w_2(k)\}_{\frac{N}{4}}, \ldots, \{w_j(k)\}_{\frac{N}{2^j}},\{w_{J}(k)\}_{\frac{N}{2^{J}}} \} \,. 
\end{equation}

The literature calls the set of coefficients \cite{veitch00b}, \cite{abry98}
\begin{equation}
\label{eq:DWT}
\left\{ \{w_1(k)\}_{\frac{N}{2}},\{w_2(k)\}_{\frac{N}{4}}, \ldots,\{w_{J}(k)\}_{\frac{N}{2^{J}}}, \{u_{J}(k)\}_{\frac{N}{2^{J}}} \right\}
\end{equation}
as the DWT of the $x(t)$ signal.

\begin{figure}[htp]
	\begin{center}
		\includegraphics[width=\textwidth,keepaspectratio]{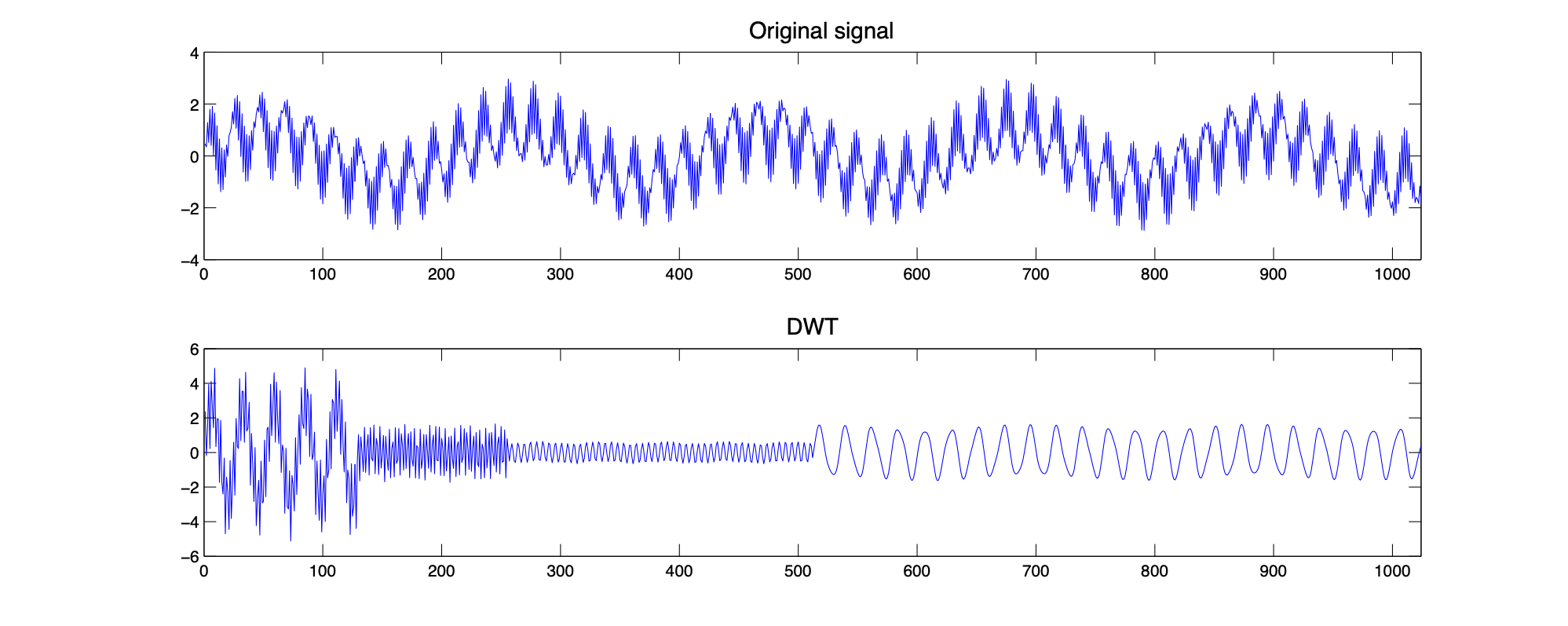}
	\end{center}
	\caption{An illustration of the 3-levels DWT of the discrete time signal 
		$x(t)= \sin{(3t)} + \sin{(0.3t)} + \sin{(0.03t)}$. The graph concatenates the sequences of the scale coefficients $\{u_3(t)\}_{128}$ and of the wavelet coefficients $\{w_3(t)\}_{128}$, $\{w_2(t)\}_{256}$ e $\{w_1(t)\}_{512}$ from left to right, i. e., the first 128 points correspond to the sequence $\{u_3(t)\}_{128}$; then follow the 128 points of the sequence $\{w_3(t)\}_{128}$, the 256 points of the sequence $\{w_2(t)\}_{256}$ and 512 points of the sequence $\{w_1(t)\}_{512}$.}
	\label{fig:wavedec-sumsin}
\end{figure}

\begin{figure}[htp]
	\begin{center}
		\includegraphics[scale = 0.35]{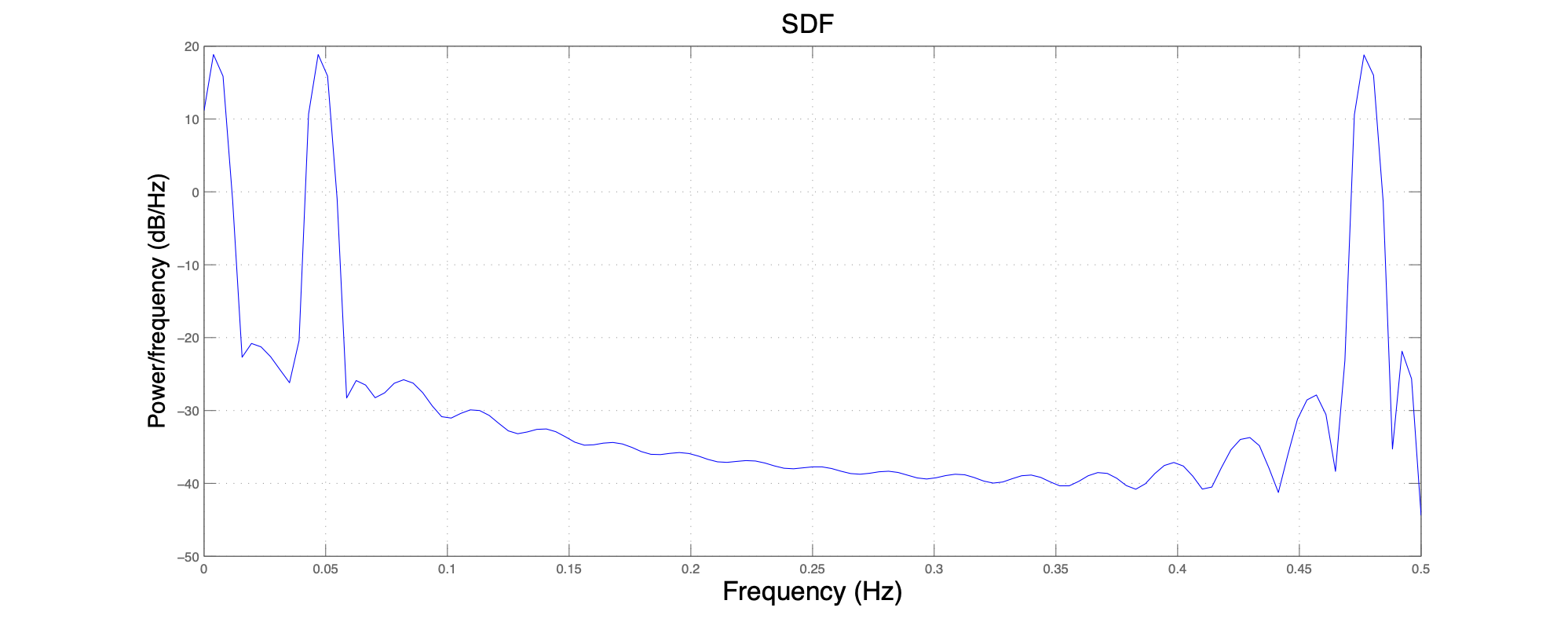}
	\end{center}
	\caption{PSD of the signal $x(t)= \sin{(3t)} + \sin{(0.3t)} + \sin{(0.03t)}$.}
	\label{fig:dep-sumsin}
\end{figure}

Fig. \ref{fig:wavedec-sumsin} illustrates a 3-level DWT (decomposition in scales $j=1,2,3$) associated to $1024$ samples of the discrete time signal $x(t)= \sin{(3t)} + \sin{(0.3t)} + \sin{(0.03t)}$, that corresponds to the superposition of 3 sinusoids in frequencies $f_1 \approx 0.004775$ cycle/sample, $f_2 \approx 0.04775$ cycle/sample and $f_3\approx 0.4775$ cycle/sample.  Fig. \ref{fig:dep-sumsin} shows the Power Spectral Density (PSD) of this signal.

The reconstruction of $x(t)$ is implemented by filtering and oversampling by a factor of $2$ (\emph{upsampling}) of the sequences (\ref{eq:seq-u}) and (\ref{eq:seq-w}), obtaining an approximation of $x(t)$ in the subspace $V_0$  

\begin{equation}
\label{def:sintese-MRA}	
\mathcal{S}_0(t) = \mathcal{S}_{J}(t)  + \mathcal{D}_1(t) + \mathcal{D}_2(t) + \dots + \mathcal{D}_{J}(t) 
\end{equation} 
or
\begin{equation}
\label{eq:IDWT}
x(t)  \approx \sum_{k} u(J,k) \phi_{J,k}(t) + \sum_{j=1}^{J}\sum_{k} w_{j,k} \psi_{j,k}(t)\,. 
\end{equation}

Eq. (\ref{eq:IDWT}) defines the Inverse Discrete Wavelet Transform (IDWT). 

We say that the function $\phi_0(t)=\phi(t)$ determines a MRA of $x(t)$ according to (\ref{def:sintese-MRA}), if it obeys the following conditions:

\begin{enumerate}
	\item intra-scale orthonormality (property \ref{subespacophi})
	\begin{equation}
	\label{eq:cond-ortog-t}
	\left\langle \phi(t-m),\phi(t-n)\right\rangle = \delta_{m,n}\,,
	\end{equation}
	in which $\delta_{m,n}$ is the Kronecker's delta ($\delta_{m,n}=1$ if $m=n$, $\delta_{m,n}=0$ for $m\neq n$). Eq.  (\ref{eq:cond-ortog-t}) imposes an orthonormality condition at scale $j=0$.
	\item unit mean
	\begin{equation}
	\label{eq:media-phi}
	\int_{-\infty}^{\infty} \phi(t)\, dt=1\,.
	\end{equation}			
	\item 
	\begin{equation}
	\label{eq:dilation}
	\frac{1}{\sqrt{2}}\phi(\frac{t}{2}) = \sum_n g_n \phi(t-n)\,,
	\end{equation}
	as several $\phi(t-k)$ fit in  $\phi(\frac{t}{2})$  (is a consequence of property (\ref{subconjuntosVj}) of the MRA).
\end{enumerate}

Equation \ref{eq:dilation} may be rewritten as

\begin{equation}
\label{eq2:dilation}
\phi(t) = \sum_n \sqrt2 g_n \phi(2t-n)\,,
\end{equation}

known as \textit{Dilation Equation}, $n \in \mathbb{Z}$.

Equations \ref{eq:dilation} and \ref{eq2:dilation} may be written, respectively, in the frequency domain as

\begin{equation}
\label{eq:escala-f}
\sqrt2 \Phi(2\nu) = G(\nu) \Phi(\nu)\,,
\end{equation}
and
\begin{equation}
\label{eq2:escala-f}
\Phi(\nu) = \frac{1}{\sqrt{2}} G(\nu) \Phi(\frac{\nu}{2})\,,
\end{equation}

in which $\Phi(\nu)$ is the FT of $\phi(t)$ and $G(\nu) = \sum_n g_n e^{-j 2 \pi \nu n}$, known as \textit{scale filter} (low-pass), represents a periodic filter in $\nu$.

As the subspace $W_{j+1}$ is orthogonal to $V_{j+1}$ and is in $V_j$, we have

\begin{equation}
\label{eq:wavelet}
\frac{1}{\sqrt{2}}\psi(\frac{t}{2}) = \sum_n h_n \phi(t-n)\,,
\end{equation}
or
\begin{equation}
\label{eq2:wavelet}
\psi(t) = \sum_n \sqrt{2} h_n \phi(2t-n)\,,
\end{equation}
that is the \textit{Wavelet Equation}.

Applying the FT to (\ref{eq:wavelet}) and (\ref{eq2:wavelet}) we get, respectively,

\begin{equation}
\label{eq:wavelet-f}
\sqrt(2) \Psi(2\nu) = H(\nu) \Phi(\nu)\,,
\end{equation}
and 
\begin{equation}
\label{eq2:wavelet-f}
\Psi(\nu) = \frac{1}{\sqrt{2}} H(\nu) \Phi(\frac{\nu}{2})\,.
\end{equation}

in which $H(\nu)$ is the \textit{wavelet filter} (high-pass).	

Rewriting (\ref{eq:<phi,psi>}) in terms of the frequency domain and using (\ref{eq:escala-f}) and (\ref{eq:wavelet-f}) results the \textit{orthogonality condition}

\begin{equation}
\label{eq:cond-ortog}
\int_{-\infty}^{\infty} G(\nu)H^{*}(\nu) |\Phi(\nu)|^2 \, d\nu = 0\,,
\end{equation}

that the filter $H$ has to obey so the family $\{\psi_{1,k}(t)\}$ is orthogonal to the family $\{\phi_{1,k}(t)\}$.

We may show that the condition \cite[p.150]{kaiser94}, \cite[p.75]{percival00}

\begin{equation}
\label{eq:QMF}
h_n = (-1)^n g_{L-1-n}\,, \quad \leftrightarrow \quad H(z)=-z^{-L+1}G(-z^{-1})\,, 
\end{equation}

in which $L$ denotes the length of a Finite Impulse Response (FIR) filter $g_n$,  $H(z)$ and $G(z)$ denote the $z$-transform of sequences $h_n$ and $g_n$, respectively,  is sufficient to (\ref{eq:cond-ortog}) to hold.  

We say that $g_n$ e $h_n$ are Quadrature Mirrored Filters (QMF) when they are related by (\ref{eq:QMF}).	

\begin{figure}[htp]
	\begin{center}
		\includegraphics[scale = 0.3]{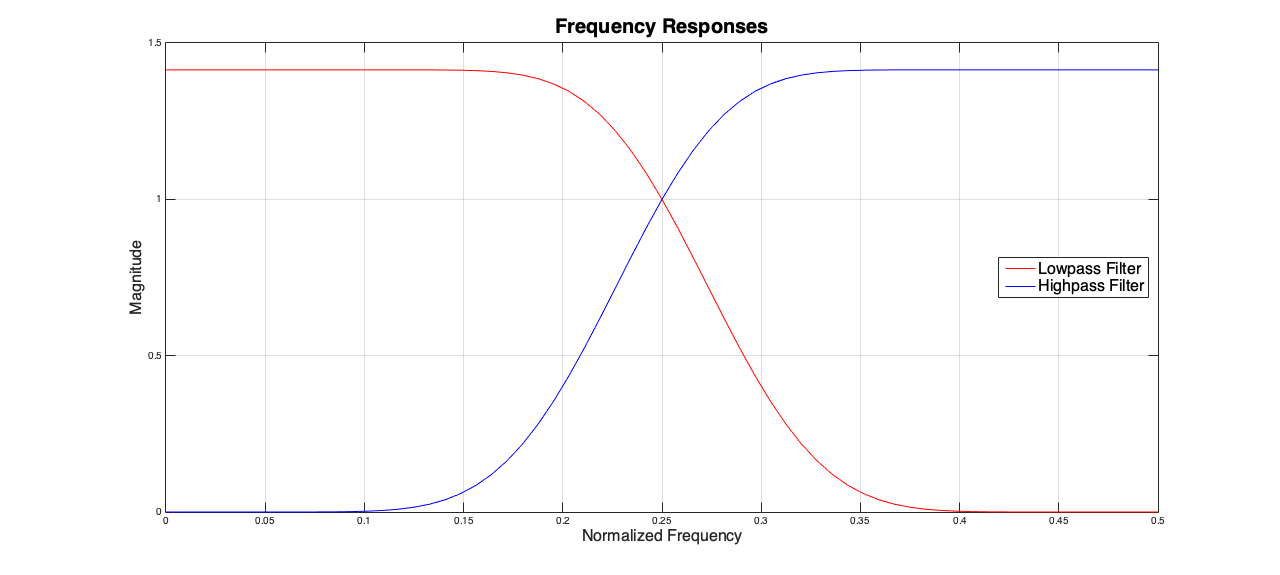}
	\end{center}
	\caption{QMF filters frequency response.}
	\label{fig:QMF}
\end{figure}

Figure \ref{fig:QMF} shows the QMF filters frequency response related to a Daubechies wavelet of order $10$ (db10).

According to (\ref{eq2:dilation}), the MRA departs from a definition (from several possible) of the scale function $\phi(t)$, that is related to the scale filter $g_n$ by (\ref{eq:dilation}). Eq. (\ref{eq:QMF}) says that the choice of a FIR-type filter $\{g_n\}$ implies a $\{h_n\}$ that is also FIR. At last, the wavelet function is determined by (\ref{eq:wavelet}). 

The scale $\phi(t)$ and wavelet $\psi(t)$ functions associated to the FIR filters $\{g_n\}$ and $\{h_n\}$ have compact support, thus offering the \emph{time resolution} functionality. The simplest scale function that satisfies (\ref{eq:cond-ortog-t}) is the characteristic function of the interval $I = [0,1)$, that corresponds to the Haar's scale function:

\begin{equation}
\label{eq:Haar-scale-function}
\phi^{(H)}(t) = \chi_{[0,1)}(t) = 
\begin{cases}
1,\,\,\,\text{if}\,\, 0 \leq t <1	\\
0,\,\,\,\text{otherwise}.
\end{cases} 
\end{equation}

In this case (Haar MRA), the associated Haar scale filter is given by

\begin{equation}
\label{eq:filtro-escala-Haar}
g_n=\{\ldots,0,g_0=1/\sqrt{2}, g_1=1/\sqrt{2},0,\ldots\}\,,
\end{equation}
the Haar wavelet filter by 
\begin{equation}
\label{eq:filtro-wav-Haar}
h_n=\{\ldots,0,h_0=g_1=1/\sqrt{2}, h_1=-g_0=-1/\sqrt{2},0,\ldots\}\,
\end{equation}
and the Haar wavelet function by
\begin{equation}
\label{eq:Haar-wav-function}
\psi^{(H)}(t) = \chi_{[0,1/2)}(t) - \chi_{[1/2,1)}(t)\,.  
\end{equation}	

\begin{figure}[htp]
	\begin{center}
		\includegraphics[width=0.9\textwidth,keepaspectratio]{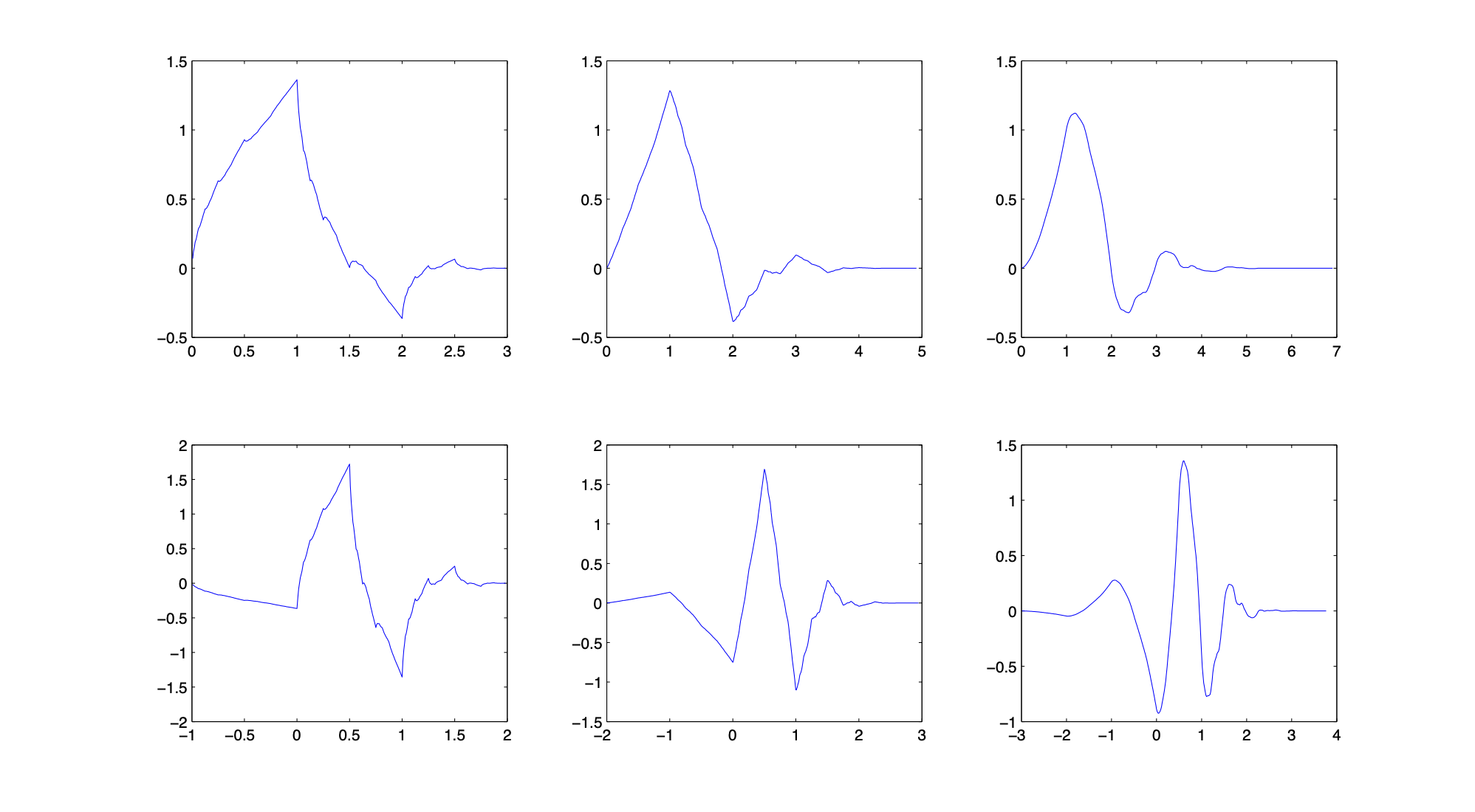}
	\end{center}
	\caption{The graphs in the lower part show the Daubechies' wavelets with $N=2,3,4$ \emph{vanishing moments}, from left to right, respectively. The corresponding scale functions are in the upper part.}
	\label{fig:exemplos-wav-Daubechies}
\end{figure}

Figure \ref{fig:exemplos-wav-Daubechies} shows the Daubechies' scale and wavelet functions with $N=2,3,4$ vanishing  moments 

\begin{equation}
\label{def:van-moments}
\int_{-\infty}^{\infty} t^m \psi(t) \, dt = 0, \,\,\, m=0,1,\ldots,N-1 \,.
\end{equation}

Ingrid Daubechies \cite{daubechies88} was the first one to propose a method for building sequences of transfer functions $\{ G^{(N)}(z) \}_{N=1,2,3,\dots}$ and $\{ H^{(N)}(z)\}_{N=1,2,3,\dots}$, in which $G^{(N)}(z)$ is associated to the low-pass FIR filter $g^{(N)}_n$ and $H^{(N)}(z)$ to the high-pass filter  $h^{(N)}_n$. The corresponding scale and wavelet functions have support in $[0,2N-1]$. The first member of the sequence is the Haar system $\phi^{(1)}=\phi^{(H)}$, $\psi^{(1)}=\psi^{(H)}$. The Daubechies' filters are generalizations of the Haar system for $N \geq 2$ \cite{kaiser94}.

We can demonstrate that \cite{mallat89}:
\begin{equation}
\label{eq1:BF}
u_{j}(n) = \sum_{k} g(k-2n) u_{j-1}(k)\, 
\end{equation}
and that
\begin{equation}
\label{eq2:BF}
w_{j}(n) = \sum_{k} h(k-2n) u_{j-1}(k)\,. 
\end{equation}	

According to (\ref{eq1:BF}) and (\ref{eq2:BF}), we can obtain the coefficients $u_j(n)$ and $w_j(n)$ from the scale coefficients $u_{j-1}(m)$ by means of decimation operation of the sequence $\{u_{j-1}(m)\}$ by a factor of 2. The decimation consists in cascading a low-pass filter $g(-m)$ (with a transfer function $\bar{G}(z)=G(1/z)$ and frequency response $G^{\ast}(f)$) or a high-pass $h(-m)$ (with transfer function $\bar{H}(z)=H(1/z)$ and frequency response $H^{\ast}(f)$) with a compressor (or decimator) by a factor of 2. Decimate a signal by a factor $D$ is the same as to reduce its sampling rate by $D$ times. 

The MRA is implemented by a low-pass and high-pass analysis filter banks $G^{\ast}(f)$ and $H^{\ast}(f)$ adequately positioned for separating the scale and wavelet coefficients sequences. This is known in the literature as the pyramid algorithm presented by Mallat \cite{mallat89}. 
Later, it is possible to rebuild the original signal using dual QMF reconstruction filter banks, low-pass $G(f)$ and high-pass $H(f)$. 

It is important to emphasize that the pyramid algorithm's complexity is $O(N)$ (assuming we want to evaluate the DWT of $N$ samples), while the direct evaluation of the DWT (that involves matrices multiplication) is $O(N^2)$ \cite{percival00}.

\begin{figure}[htp]
	\centering
	\includegraphics[scale = 0.7]{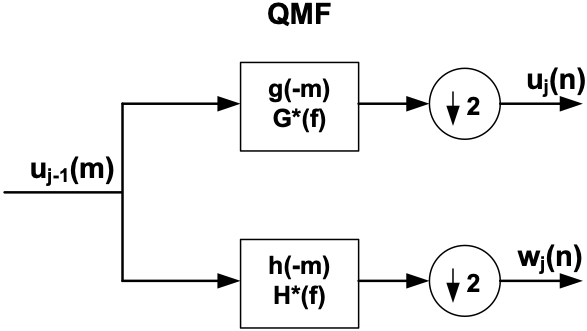}
	\caption{QMF analysis filter banks  $G^{\ast}(f)$ (low-pass) and $H^{\ast}(f)$ (high-pass) with decimation (\emph{downsampling}) by a factor of 2.}
	\label{fig:QMF-Bank-Analise}
\end{figure}

\begin{figure}[htp]
	\centering
	\includegraphics[scale = 0.7]{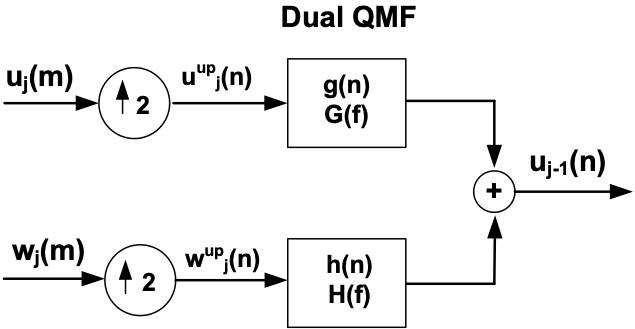}
	\caption{QMF reconstruction filter banks with interpolation (\emph{upsampling}) by a factor of 2. Observe that are used dual low-pass and high-pass filters, $G(f)$ and $H(f)$.}
	\label{fig:QMF-Bank-Recon}
\end{figure}

Figure \ref{fig:QMF-Bank-Analise} shows the QMF analysis filter banks  $G^{\ast}(f)$ (low-pass) and $H^{\ast}(f)$ (high-pass) with decimation (\emph{downsampling}) by a factor of 2.  Figure \ref{fig:QMF-Bank-Recon} shows the QMF reconstruction filter banks with interpolation (\emph{upsampling}) by a factor of 2. Observe that are used dual low-pass and high-pass filters, $G(f)$ and $H(f)$.

\begin{figure}[htp]
	\centering
	\includegraphics[scale = 0.7]{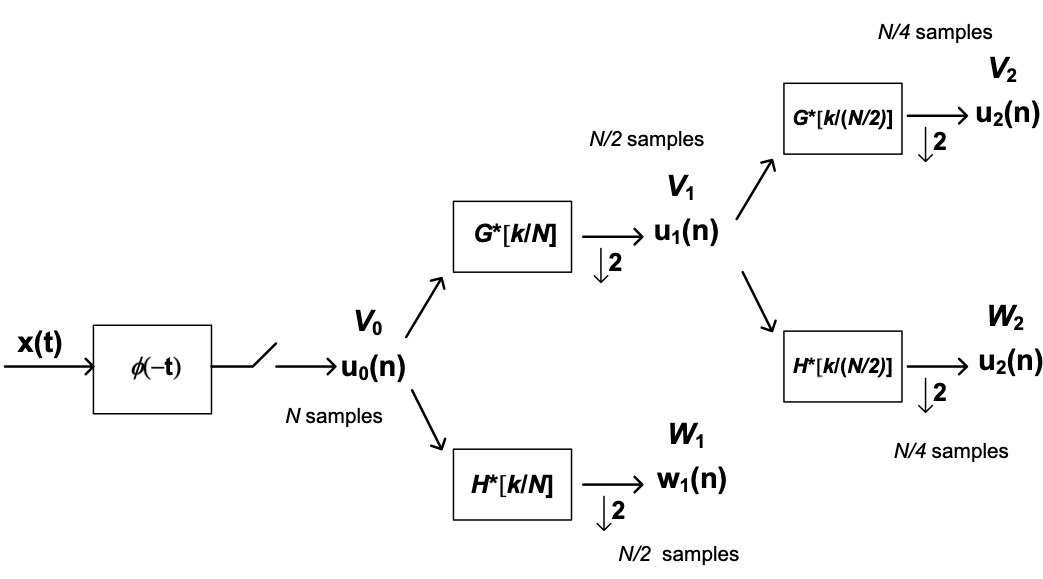}	
	\caption{Flow diagram that shows the initial projection of a signal $x(t)$ on $V_{0}$ followed by the decomposition in  $W_1$, $W_2$ and $V_2$.}
	\label{fig:MRA}
\end{figure}

Figure \ref{fig:MRA} presents the flow diagram that shows the initial projection of a signal $x(t)$ on $V_{0}$ followed by the decomposition in  $W_1$, $W_2$ and $V_2$.

\begin{figure}[htp]
	\centering
	\includegraphics[scale = 0.7]{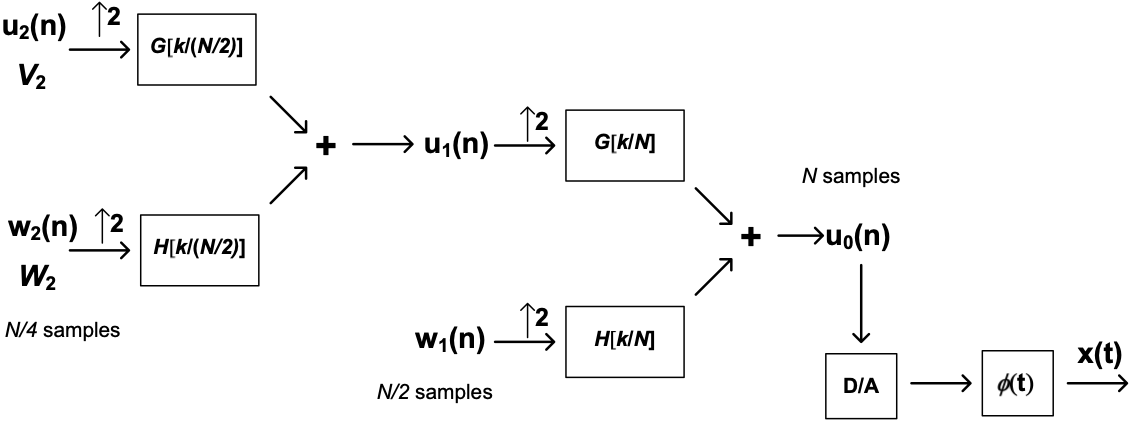}
	\caption{Flow diagram that illustrates the approximate synthesis of $x(t)$ from $W_1$, $W_2$ and $V_2$.}%
	\label{fig:recon}
\end{figure}

Figure \ref{fig:recon} shows the flow diagram that illustrates the approximate synthesis of $x(t)$ from $W_1$, $W_2$ and $V_2$.

\begin{figure}[htp]
	\begin{center}
		\includegraphics[scale = 0.8]{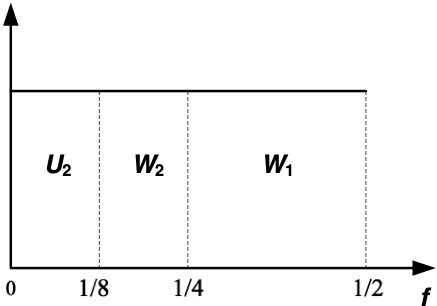}
		\caption{The spectrum $U_0(f)$ of the signal $u_0(n)$ is subdivided in three frequency bands (that cover two octaves): $0 \leq f < 1/8$, $1/8 \leq f < 1/4$ and $1/4 \leq f \leq 1/2$.}
		\label{fig:subband}
	\end{center}
\end{figure}

Figure \ref{fig:subband} presents a block diagram that shows that the DWT works as a sub-bands codification scheme. The spectrum $U_0(f)$ of the signal $u_0(n)$ is subdivided in three frequency bands (that cover two octaves): $0 \leq f < 1/8$, $1/8 \leq f < 1/4$ and $1/4 \leq f \leq 1/2$.

\section{Experimental Results}\label{sec:results}

In section \ref{sec:wavelets}, we described how the DWT can be applied to a signal $\rvx(t)$. However, the purpose of this study requires that we think of $\rvx(t)$ as a realization of a stochastic process $\{\rvx(t)\}$, so that we can be able to realize a statistical assessment of the change point. The reader interested in the definition of a stochastic process can consult the reference \cite{lima2020a} for more information.

\begin{figure}[ht] 
	\begin{center}
		\includegraphics[scale = 0.40]{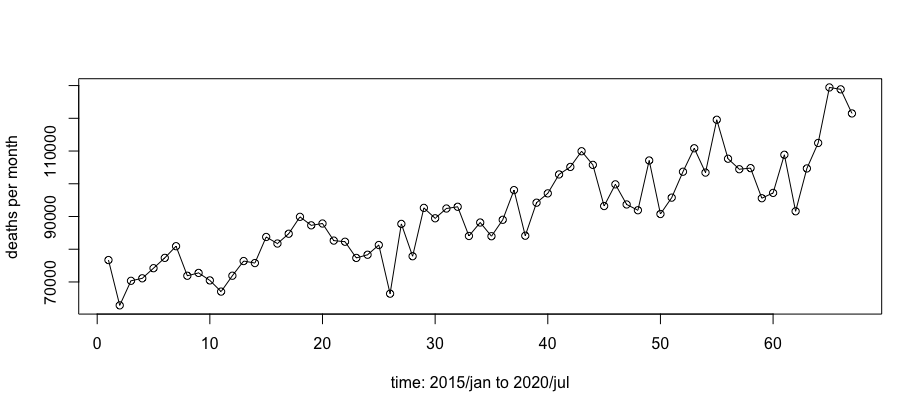} 
	\end{center}
	\caption{Historical series in Brazil from January 2015 to July 2020.}
	\label{fig:serie-original}
\end{figure}

Figure \ref{fig:serie-original} shows the historical series in Brazil from January 2015 to July 2020 ($67$ samples).

In the paper \cite{lima2020a}, we showed that the signal of Fig. \ref{fig:serie-original} obeys a model of the type \cite{shumway2006}[p. 58]

\begin{equation}
\label{eq:linear-trend-model}
\vax(t)  = \mu(t) + \vay(t),
\end{equation}

where $\vax(t)$ are the observations, $\vay(t)$ is a stationary process, and $\mu(t) $ denotes a linear trend given by the regression model

\begin{equation}
\label{eq:linearmodel}
	\mu(t)  = \beta_0 + \beta_1 t
\end{equation}

in which $\beta_0$ and $\beta_1$ are the intercept and the slope parameters. We also estimated  an autoregressive model of order $11$ ($\text{AR}(11)$) for $\vay(t)$. An $\text{AR}(11)$ is said to exhibit short-range dependence, as the Power Spectral Density (PSD) of the signal does not have $1/f$ behavior for frequencies near to zero (Long Range Dependence (LRD)). 

In \cite{lima2020a}, we conclude that there are no change points in the slope of the historical series and in the mean of the series that corresponds to the first difference of the historical series.

\begin{figure}[ht] 
	\begin{center}
		\includegraphics[scale = 0.40]{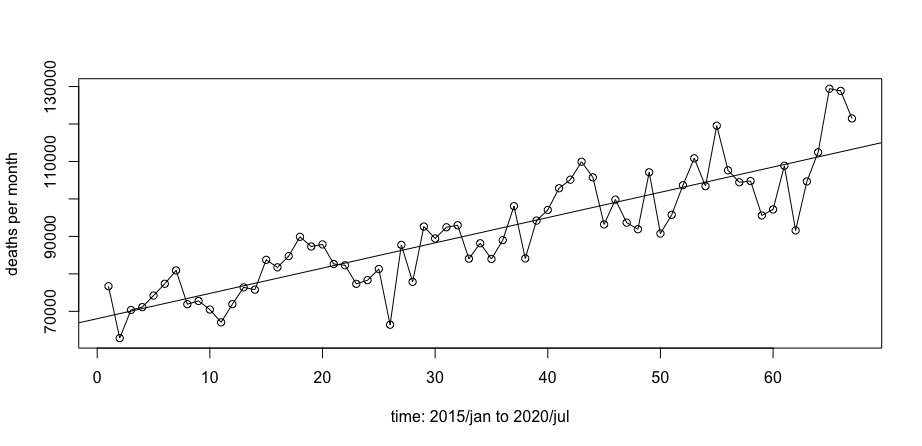} 
	\end{center}
	\caption{Historical series with the superimposed linear regression model.}
	\label{fig:serie-original-fit}
\end{figure}

Table \ref{tab:estimated-model-mu} shows the estimated coeffcientes for (\ref{eq:linearmodel}) and its $p$-values. Figure \ref{fig:serie-original-fit} shows the historical series with the superimposed linear regression model.

\begin{table}[htp]
	\caption{Estimated model for $\mu_t$.}
	\centering
	\begin{tabular}{c c c c c c c c c} \hline\hline  
		$\widehat{\beta_0}$		 & $p$-value of  $\widehat{\beta_0}$  &  $\widehat{\beta_1}$       & $p$-value of  $\widehat{\beta_1}$ \\ \hline
		$68,036.6$                        &   $< 2.2e^{-16}$                                          &  674.7                                   &  $< 2.2e^{-16}$				 \\ \hline\hline
	\end{tabular}
	\label{tab:estimated-model-mu}
\end{table}

As in \cite{lima2020a}, the first step in wavelet analysis requires that the deterministic trend of the historical series be removed, in order to perform the time-frequency domain analysis \cite{shumway2006}[p. 58]. There are two alternatives: remove the line estimated by the regression or take the first difference in the series. As our goal is to coerce the data to (a possible) stationarity, then differencing may be more appropriate \cite{shumway2006}[p. 61].  The first two samples of the historical series were discarded so that the series corresponding to the first difference has 64 points, that is, $2^6$ points, which is appropriate for a wavelet analysis with the DWT. 

The first difference can be denote as

\begin{equation}
\label{eq:first-diff}
\Delta \rvx(t) = \rvx(t) - \rvx(t-1) = \var(t).
\end{equation}

\begin{figure}
	\begin{center}
	\includegraphics[scale = 0.40]{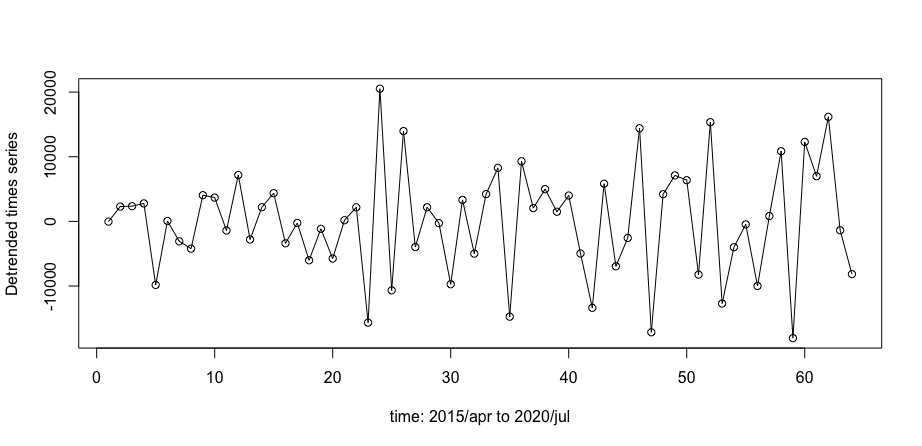}
	\end{center}
	\caption{first difference series.}
	\label{fig:first-diff}
\end{figure}

Figure \ref{fig:first-diff} shows the series $\var(t)$ (we also demeaned it).

Now, as far as a qualitative analysis is concerned, we have to focus our  wavelet (graphical) analysis on the most refined scales of the time-frequency doman, where possible changes in variance can be localized. This is the intuition behind the statistical test of homogeneity of variance proposed by Percival and Walden \cite[p. 380]{percival00}. The null hypothesis at level $j$ of the DWT is given by

\begin{equation}
\label{eq:test-variance}
	H_0:\quad\text{var}\{w_j(L^{'}_j)\} = \text{var}\{w_j(L^{'}_j + 1)\} = \ldots = \text{var}\{w_j(N_j-1)\},
\end{equation}

where we assume that $\var(t)$ has $N^{'}_j$ nonboundary wavelet coefficients $w_j(L^{'}_j), w_j(L^{'}_j + 1), \ldots, w_j(N_j-1)$, in which $N^{'}_j=N_j-L^{'}_j$.

\begin{figure}[htp]
	\begin{center}
		\includegraphics[scale = 0.5]{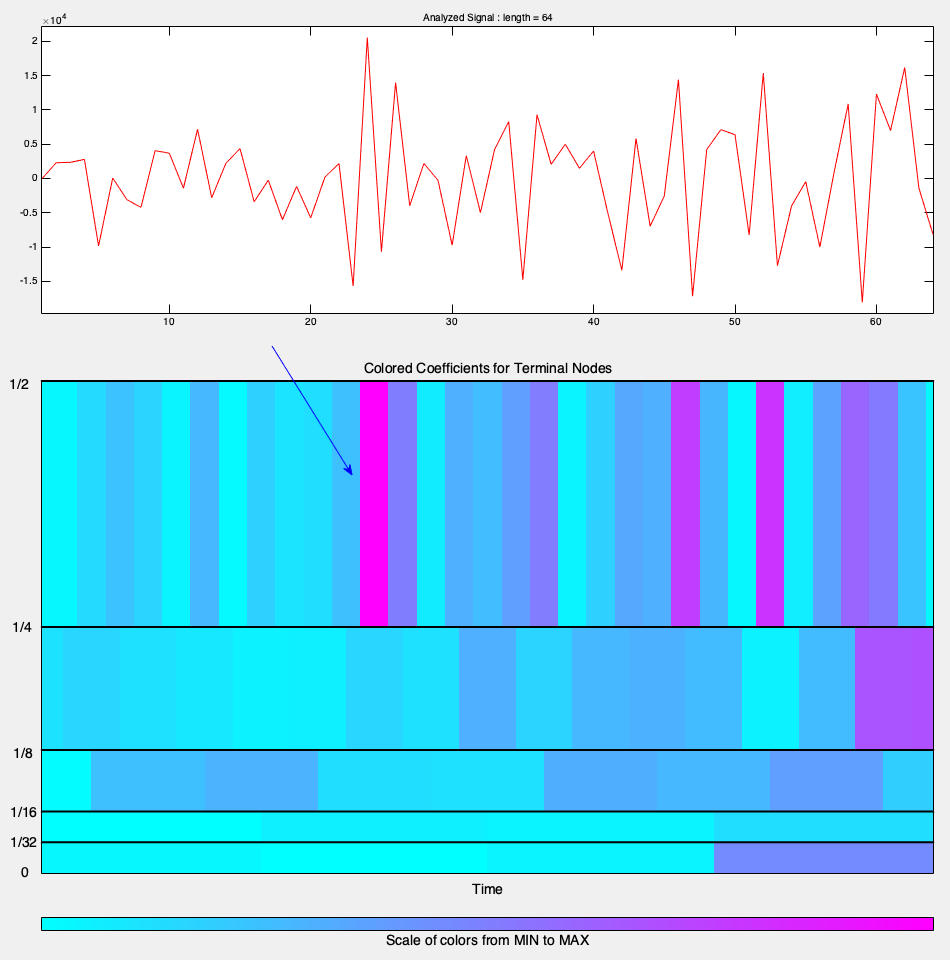}	
	\end{center}
	\caption{LA(8) DWT of signal $\var_t$ (four levels).}
	\label{fig:dwt-LA8-toolbox-matlab}
\end{figure}

\begin{figure}[htp]
	\begin{center}
		\includegraphics[scale = 0.35]{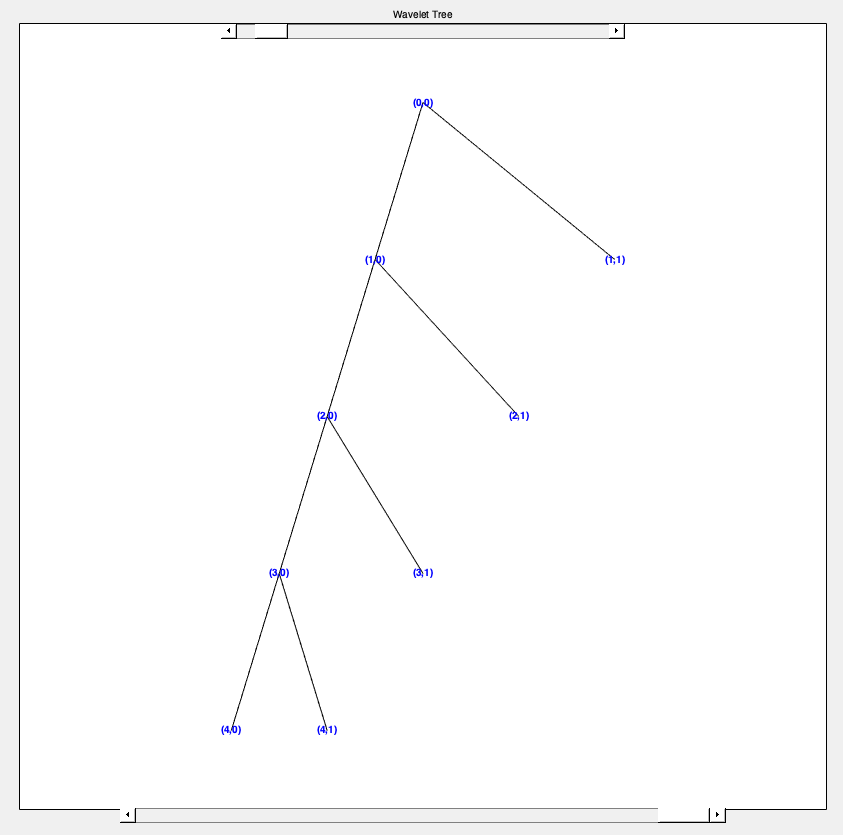}	
	\end{center}
	\caption{Wavelet tree of the LA(8) DWT (four levels).}
	\label{fig:wavelet-tree}
\end{figure}

Figs. \ref{fig:dwt-LA8-toolbox-matlab} and \ref{fig:wavelet-tree} show the LA(8) DWT of signal $\var_t$ and its wavelet tree, respectively. The wavelet coefficients of the node $(1,1)$ of the tree in Fig.\ref{fig:wavelet-tree} are between the frequencies $1/4$ and $1/2$. The change point is indicated by the blue arrow ($t=22$) in Fig. \ref{fig:dwt-LA8-toolbox-matlab}. Note that this point is localized just before the maximum value of the set of coefficients $\{w_1(k)\}_{\frac{64}{2}}$. The LA(8) wavelet has an approximate linear phase, which facilitates the alignment of the wavelet coefficients with the signal of interest. Thus, this strategy is often a good choice when we want to detect a change point in the signal \cite[p. 136]{percival00}. 

Here is a noteworthy example of an important advantage of wavelet analysis over Fourier. Should the same signal had been analyzed by the FT, we would not have been able to detect the instant of the change point, whereas it is clearly observable here.

\begin{figure}[htp]
	\begin{center}
		\includegraphics[scale = 0.40]{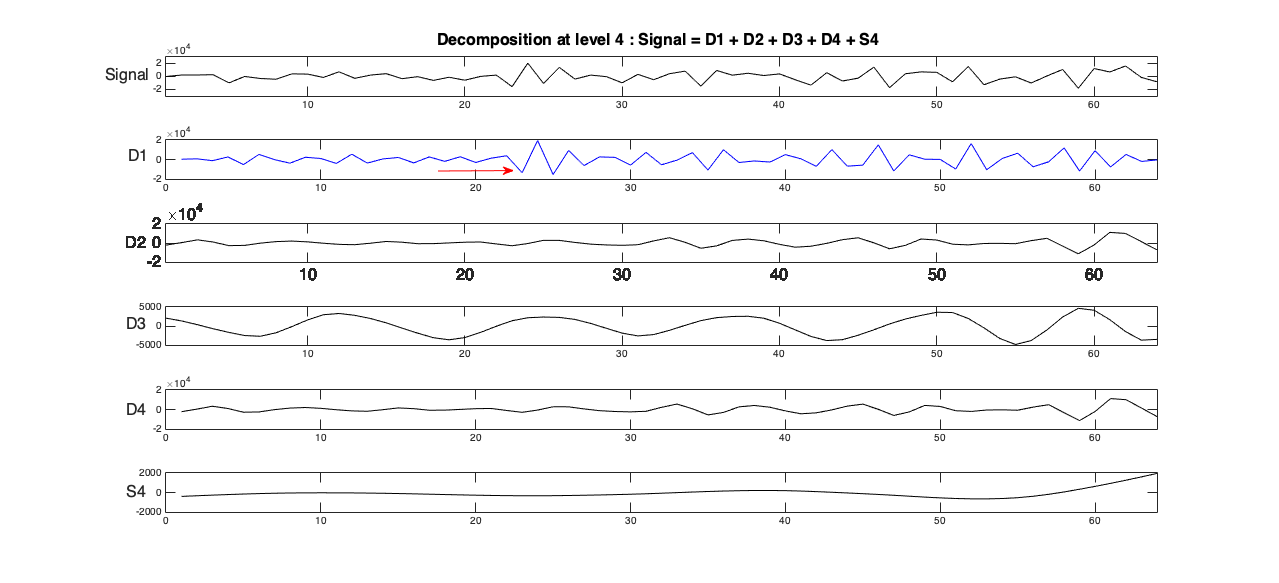}	
	\end{center}
	\caption{LA(8) MRA analysis of signal $\var_t$ (four levels).}
	\label{fig:LA8-MRA-diff1}
\end{figure}

\begin{figure}[htp]
	\begin{center}
		\includegraphics[scale = 0.40]{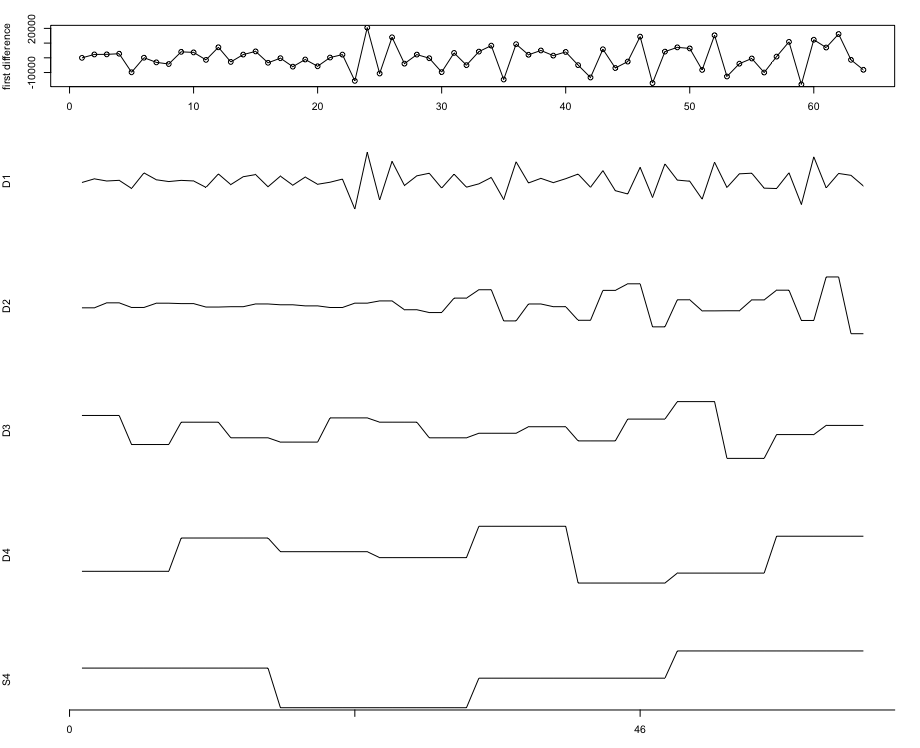}	
    \end{center}
	\caption{Haar MRA analysis of signal $\var_t$ (four levels).}
	\label{fig:Haar-MRA-diff1}
\end{figure}

Now, let us compare the plot of the raw data along with the level-one wavelet detail signal (D1) in Fig. \ref{fig:LA8-MRA-diff1} (LA(8) MRA), which indicates the existence of a  change in variance around $t=22$ (2017/jan), see the red arrow. Remember that the detail signal D1 is located in the frequency band $\left [\frac{1}{4},\frac{1}{2}\right]$. Fig. \ref{fig:Haar-MRA-diff1} illustrates the Haar MRA. Note that we can also see change in variance around $t=22$ (2017/jan) in the level-one wavelet detail signal.

Now we can confirme our qualitative analysis using the function \texttt{wvarchg()} (MATLAB\textcopyright$\,$R2015a Wavelet Toolbox), which calculates the optimal positioning and (potentially) number of changepoints. As the signal $\var(t)$ is SRD, there is no need to be concerned with the question of sampling of the variance of the wavelet coefficients, which is problematic for long memory process \cite[p. 380]{percival00}.

The test confirms that there is a change point in variance in $t=22$. 

\section{Conclusions and Future Work}\label{sec:conclusions}

In this paper, we presented an overview of the Continuos and Discrete Wavelet Transforms, and a wavelet analysis of the historical series of the total number of deaths per month in Brazil since 2015. Our preliminary results indicate that: 

\begin{itemize}
	\item the variance of the signal changed in january of 2017; however, this change point occurred before the outbreak of the COVID-19 pandemic; and
	\item there is no evidence that COVID-19 provoked a change in the stochastic process that generates the historical series, as there are no change points in this signal after the outbreak of COVID-19 in Brazil, which happened in the beginning of march, 2020.
\end{itemize}

There is no doubt that COVID-19 has caused the deaths of many people around the world. Many who survived will have to live with sequelae in the brain, kidneys, lungs and heart. In addition, the world population is being subjected to a high degree of stress, due to fear of the new coronavirus, economic/financial problems, etc.

However, our results suggest that COVID-19 \textit{did not cause any change point in the brazilian series of the total number of deaths per month in Brazil since 2015 so far}. What would be a plausible explanation for this strange result? Answering this question requires a multidisciplinary approach, as it involves several areas of knowledge such as medicine, signal processing, statistics etc.

Nevertheless, researchers in the area must continue to monitor the behavior of the historical series in Brazil.

In future work, we suggest the application of the Maximal Overlap Discrete Wavelet Transform (MODWT) and the Discrete Wavelet Packet Transform (DWPT) \cite{percival00} to the brazilian historical series. We would also like to analyze the series from other countries affected by COVID-19.

\bibliographystyle{unsrt}
\bibliography{references}

\begin{thebibliography}{10}

\bibitem{minsaude}
{Brazilian Federal Ministry of Health (Ministério da Saúde do Brasil)}.
\newblock {COVID-19 no Brasil}, 2020.

\bibitem{WHO}
{World Health Organization}.
\newblock {Coronavirus disease (COVID-19) pandemic}, 2020.

\bibitem{JHU}
{Johns Hopkins University}.
\newblock {COVID-19 Dashboard by the Center for Systems Science and Engineering
  (CSSE) at Johns Hopkins University (JHU)}, 2020.

\bibitem{percival00}
D.~B. Percival and A.~T. Walden.
\newblock {\em Wavelet Methods for Time Series Analysis}.
\newblock Cambridge University Press, 2000.

\bibitem{chen2000}
J.~Chen and A.~K. Gupta.
\newblock {\em Parametric statistical change point analysis}.
\newblock Birkhauser, 2000.

\bibitem{killick2013}
R.~Killick, I.~A. Eckley, and P.~Jonathan.
\newblock A wavelet-based approach for detecting changes in second order
  structure within nonstationary time series.
\newblock {\em Electron. J. Statist.}, 7:1167--1183, 2013.

\bibitem{lima2013}
Alexandre~Barbosa de~Lima and José~Roberto de~Almeida~Amazonas.
\newblock {\em Internet Teletraffic Modeling and Estimation}.
\newblock Gistrup: Rivers Publishers, 2013.

\bibitem{database}
{Civil Registry Offices of Brazil (Cartórios de Registro Civil do Brasil)}.
\newblock {Transparency Portal (Portal da Transparência)}, 2020.

\bibitem{cnj}
CNJ.
\newblock {Conselho Nacional de Justiça}, 2020.

\bibitem{R}
{R Core Team}.
\newblock {The R Project for Statistical Computing}, 2020.

\bibitem{GitHub}
Alexandre~B. de~Lima.
\newblock {Code and Data}, 2020.

\bibitem{gabor46}
D.~Gabor.
\newblock Theory of communication.
\newblock {\em J. Inst. Eletr. Eng.}, 93(III):429--457, 1946.

\bibitem{kaiser94}
G.~Kaiser.
\newblock {\em A Friendly Guide to Wavelets}.
\newblock Birkh\"auser, Boston, Mass., 1994.

\bibitem{whitcher2001}
Ramazan Gen\c{c}ay, Faruk Sel\c{c}uk, and Brandon Whitcher.
\newblock {\em An Introduction to Wavelets and Other Filtering Methods in
  Finance and Economics}.
\newblock Academic Press, 2001.

\bibitem{daubechies88}
Ingrid Daubechies.
\newblock Orthonormal bases of compactly supported wavelets.
\newblock {\em Comm. Pure Appl. Math.}, 41:909--996, 1988.

\bibitem{morlet84}
A.~Grossmann and J.~Morlet.
\newblock Decomposition of hardy functions into square integrable wavelets of
  constant shape.
\newblock {\em SIAM J. Math.}, 15:723--736, 1984.

\bibitem{mallat89}
S.~G. Mallat.
\newblock A theory for multiresolution signal decomposition: The wavelet
  representation.
\newblock {\em IEEE Transactions on Pattern Analysis and Machine Intelligence},
  11:674--693, 1989.

\bibitem{mallat89b}
S.~G. Mallat.
\newblock Multiresolution approximations and wavelet orthonormal bases of
  $l^2(\mathbb{R})$.
\newblock {\em Transactions of the American Mathematical Society}, 315:69--87,
  1989.

\bibitem{mallat89c}
S.~G. Mallat.
\newblock Multifrequency channel decompositions of images and wavelet models.
\newblock {\em IEEE Transactions on Acoustics, Speech, and Signal Processing},
  37:2091--2110, 1989.

\bibitem{daubechies92}
I.~Daubechies.
\newblock {\em Ten Lectures on Wavelets}.
\newblock SIAM, Philadelphia, 1992.

\bibitem{veitch00b}
D.~Veitch, M.~S. Taqqu, and P.~Abry.
\newblock Meaningful \uppercase{MRA} initialization for discrete time series.
\newblock {\em Signal Processing}, 80:1971--1983, 2000.

\bibitem{mallat99}
S.~Mallat.
\newblock {\em A Wavelet Tour of Signal Processing}.
\newblock Academic Press, second edition, 1999.

\bibitem{abry98}
P.~Abry and D.~Veitch.
\newblock Wavelet analysis of long-range dependent traffic.
\newblock {\em IEEE Transactions on Information Theory}, 4(1):2--15, 1998.

\bibitem{lima2020a}
Alexandre barbosa~de Lima.
\newblock {An exploratory time series analysis of total deaths per month in
  Brazil since 2015}.
\newblock 2020.

\bibitem{shumway2006}
Robert~H. Shumway and David~S. Stofer.
\newblock {\em Time Series Analysis and Its Applications with R Examples}.
\newblock Springer, 2nd edition, 2006.

\end{thebibliography}

\end{document}